\documentclass[letterpaper,twoside,twocolumn,english,prl]{revtex4}
\usepackage[T1]{fontenc}
\usepackage[latin9]{inputenc}
\usepackage{amsmath,xcolor}
\usepackage{esint}
\usepackage{graphicx}
\usepackage{placeins}
\usepackage{float}
\usepackage{subfigure}

\makeatletter

\@ifundefined{textcolor}{}
{%
 \definecolor{BLACK}{gray}{0}
 \definecolor{WHITE}{gray}{1}
 \definecolor{RED}{rgb}{1,0,0}
 \definecolor{GREEN}{rgb}{0,1,0}
 \definecolor{BLUE}{rgb}{0,0,1}
 \definecolor{CYAN}{cmyk}{1,0,0,0}
 \definecolor{MAGENTA}{cmyk}{0,1,0,0}
 \definecolor{YELLOW}{cmyk}{0,0,1,0}
}

\usepackage{babel}

\makeatother

\usepackage{babel}
\begin{document}

\title{Dynamics of dense hard sphere colloidal systems: a numerical analysis}

\author{Paolo Sibani$^{1}$ and Carsten Svaneborg$^{1}$}

\affiliation{$^{1}$Department of Physics, Chemistry, and Pharmacy, University of Southern Denmark, Campusvej 55, DK5230, Odense M, Denmark}
\begin{abstract}
 The applicability to dense  hard sphere colloidal suspensions   of a general coarse-graining approach 
called   Record Dynamics (RD) is tested by  extensive molecular dynamics simulations.
 We reproduce known results as 
  logarithmic diffusion and the logarithmic decay of the
average potential energy per particle. We provide quantitative measures
for the cage size and identify the displacements of single particles
corresponding to cage breakings.
We then partition the system into spatial domains. Within each domain, a subset of intermittent events
called quakes is shown to constitute a log-Poisson process, as predicted by Record Dynamics.
Specifically, these events are shown to be  statistically independent
and Poisson distributed with an average depending on the logarithm of time.
Finally, we discuss the  nature of the dynamical barriers surmounted by quakes
and   link  RD to the phenomenology of aging hard sphere colloids.
\end{abstract}
\maketitle

\section{Introduction\label{sec:Introduction}}
Hard sphere colloidal  suspensions (HSC) are a paradigmatic and intensively
 investigated    complex system~\cite{Weeks02,Courtland03,ElMasri05,Cianci06,Li10,ElMasri10,Valeriani11,Hunter12},
 featuring  two different dynamical  regimes~\cite{Hunter12}: 
a time translationally invariant diffusive regime below a critical volume
fraction and, above it, an aging regime, where time homogeneity is lost.
Here, the particle mean square displacement (MSD) grows 
at a decelerating rate
through all experimentally accessible time scales.

Coarse-graining  non-equilibrium  processes as the above usually requires
the identification of  the  degrees of freedom and/or key dynamical events which  
control  the system evolution. A natural starting point in 
glassy dynamics  is  spatial heterogeneity,
the fact that only a small fraction of the system's particle 
is  dynamically active in any observational time interval~\cite{Chaudhuri07}.
The dichotomy between active and inactive (or `fast' and `slow') particles
 is demonstrated in~\cite{Chaudhuri07}
by direct trajectory inspections and by measuring   the self part of the
Van Hove distribution, where the fast particles produce an exponential tail.
In glass-formers~\cite{Pastore14} the two 
 different types of  motion are 
reversible `in-cage rattlings',  where a particle  moves reversibly within  
the small region bounded by its neighbors, and `cage-breakings', 
where it performs larger displacements which alter its neighborhood relations.
Cage rattlings are overwhelmingly the most frequent events, but  not
being associated to a  net translation,
 diffusive spreading of particles in glass-formers~\cite{Pastore14} and  diluted colloidal systems is caused
 by the much rarer cage breakings. Hence, these events  carry 
 the evolution of the system configurations and are the key
 to coarse-grain their dynamics.

%
%
Cage breakings have been  modelled~\cite{Ciamarra16,Pastore16} 
using  a Continuous Time Random Walk (CTRW)~\cite{Scher75}, a popular coarse-graining device recently
criticized in~\cite{Sibani13,Boettcher18}.
Other approaches   to coarse-graining  are  e.g. Synergetics~\cite{Haken77},
Self Organized Criticality~\cite{Bak96} and Record Dynamics (RD)~\cite{Sibani03,Anderson04,Sibani14}.
The latter  posits that   the 
 decelerating evolution of a variety of   complex 
 dynamical systems, aka `aging', is controlled by  increasingly rare  non-equilibrium events
 termed  `quakes'. 
In RD, the physical appearance of a quake is
  system dependent~\cite{Sibani99a,Oliveira05,Sibani06a,Boettcher11,Sibani16,Sibani18}, 
 but quaking is in all cases described 
 as a log-Poisson process, i.e. a Poisson process  where the  number  of events
 expected between times $t_{\rm w}<t $ and $t$ is proportional to the `log-waiting time'
 $\ln t - \ln t_{\rm w} = \ln(t/t_{\rm w})$.
 
\begin{figure}[h!]
\vspace{-3cm}
\includegraphics[width=\linewidth]{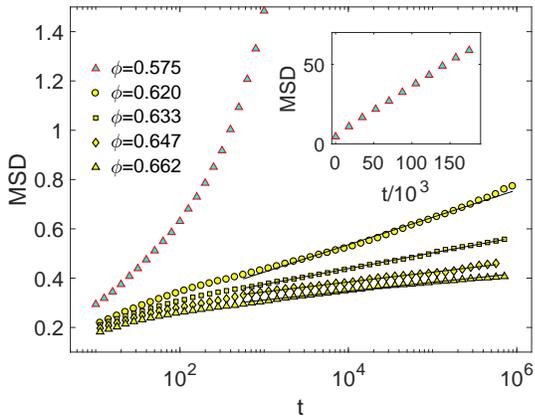}
\vspace{-3cm}
\caption{The mean square particle displacements  for several volume fractions
are plotted with a logarithmic time scale.
The insert highlights the   standard diffusive  behavior of the $\phi=0.576$ system.
}
\label{fig:MSD}
\end{figure}

Diffusion in HSC has attracted both 
experimental~\cite{Weeks02,Courtland03,ElMasri05,Cianci06,Li10} 
and computational~\cite{ElMasri10,Valeriani11}  work,
but `logarithmic diffusion' in dense HSC  is not yet  widely acknowledged.
That the particles' Mean Square Displacement (MSD)
grows with the logarithm of time was observed~\cite{Boettcher11}  in  a re-analysis  of 3D confocal
 microscopy data by Courtland et al.~\cite{Courtland03}, a behavior fully confirmed 
by the present data (see Fig.~\ref{fig:MSD}).

Accompanied by theoretical analyses and model calculations~\cite{Boettcher11,Becker14}, 
these observations promote RD as a coarse-grained description of aging dynamics in  HSC systems.
The validity of the RD description was further supported by  the  analysis~\cite{Robe16}
of experimental 2D data provided by  Yunker et al.~\cite{Yunker09}
 and by recent  Molecular Dynamics (MD) study 
of a 2D colloidal system~\cite{Robe18} which confirms two related RD
predictions: \emph{i)} the rate of quakes is
inversely proportional to time and \emph{ii)} the particles' MSD
grows logarithmically in time. These results are presently extended  by explicitly
showing the Poisson nature of the quake statistics in a 3D dense colloidal suspension..

\begin{figure}[h!]
\vspace{-3cm}
\includegraphics[width=\linewidth]{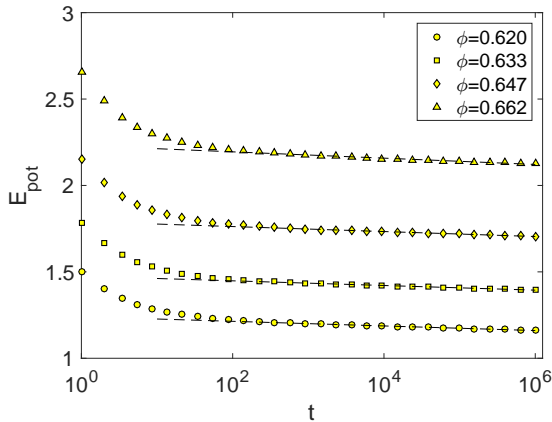}
\vspace{-3cm}
\caption{The average potential energy per particle is plotted vs. time for four different volume fractions. After an initial transient the decay appears to be linear in the logarithm of time as indicated by the fits (denoted by staggered lines).
}
\label{fig:epot}
\end{figure}

 \begin{figure*}[t!]
\vspace{-3cm}
$
\begin{array}{lr}
\includegraphics[width=.45\linewidth]{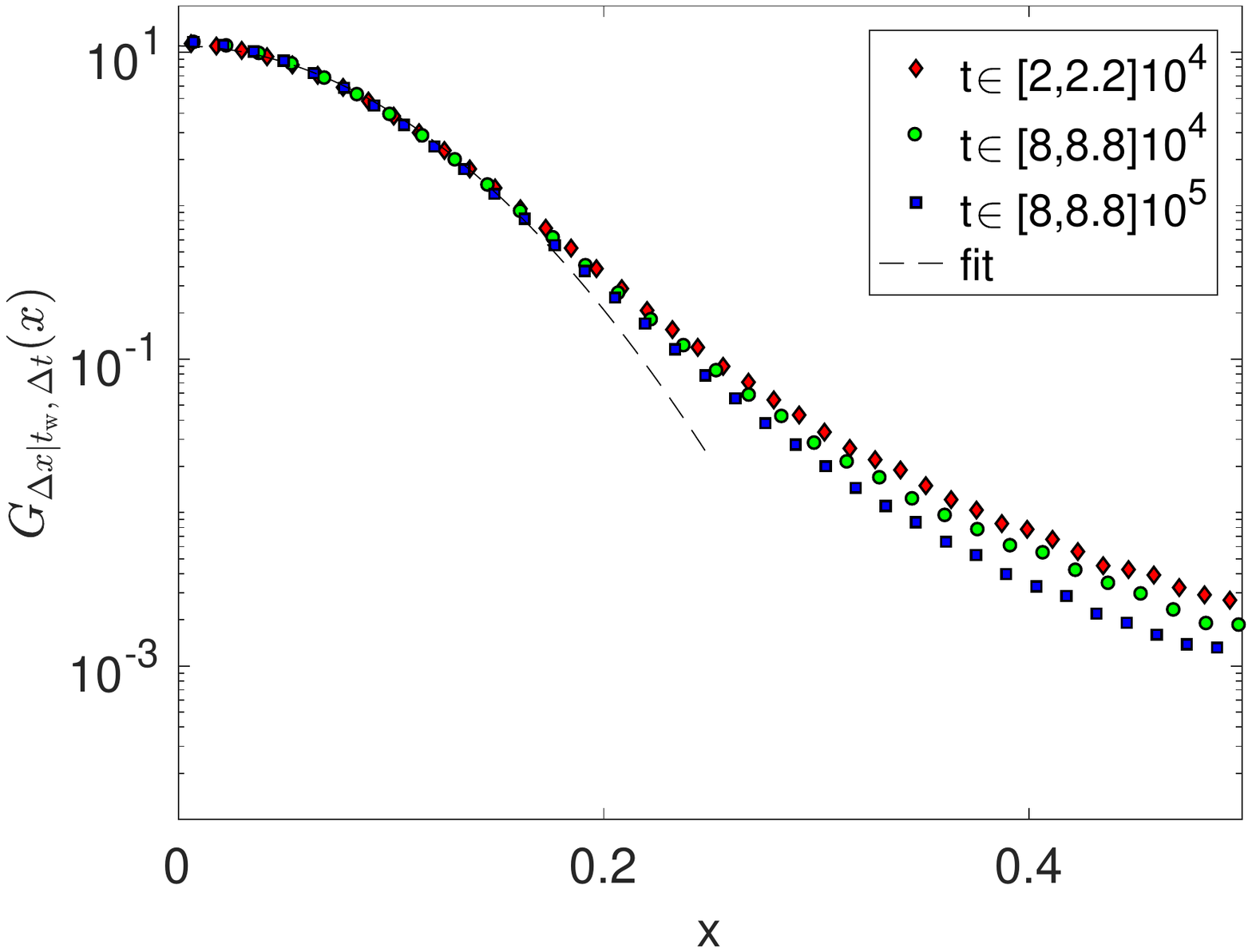} &
\includegraphics[width=.45\linewidth]{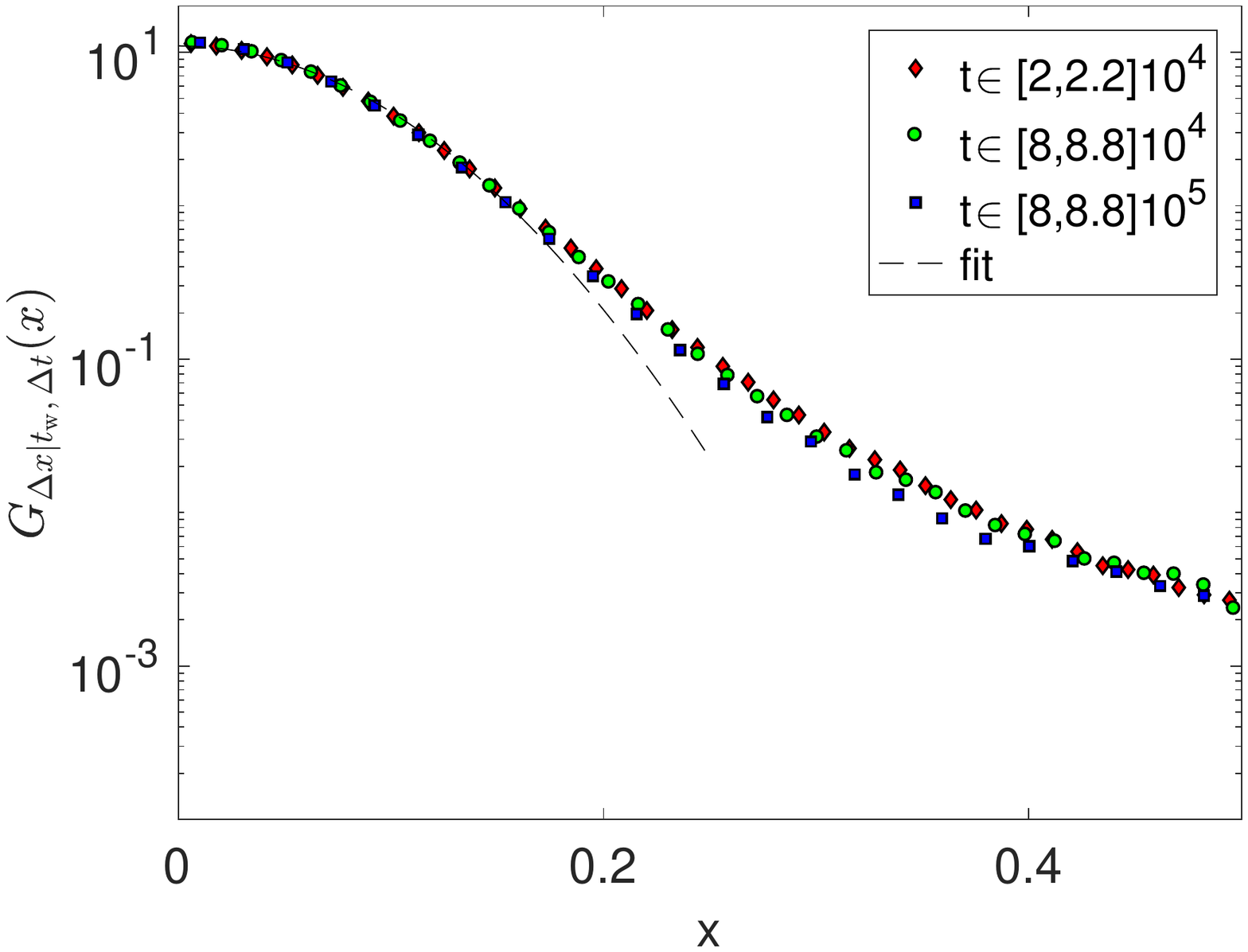}\\
\vspace{-0cm}
\end{array}
$
\vspace{-3cm}
\caption{For three different ages, $t_{\rm w}=2\cdot 10^4, 8\cdot 10^4$ and $8\cdot 10^5$, the PDF 
of  $\Delta x$,  a one dimensional  particle displacement sampled over a time interval $\Delta t \ll t_{\rm w}$,
is plotted with a logarithmic ordinate.
In both panels, the staggered line is   a fit to a Gaussian of mean $\mu_G=0$ 
and standard deviation $\sigma_G=0.05\sigma$, where $\sigma$ is the average particle diameter.
%
%
Left hand panel: the same  time interval $\Delta t=100\tau$ is used for all three values of $t_{\rm w}$.
Right hand  panel: time intervals $\Delta t=100, 400, 4000\tau$ growing proportional to the system age are used. The volume fraction of this system is $\phi=0.620$.
}
\label{fig:intermittency}
\end{figure*}
\begin{figure}[h]
\vspace{-3cm}
\includegraphics[width=.95\linewidth]{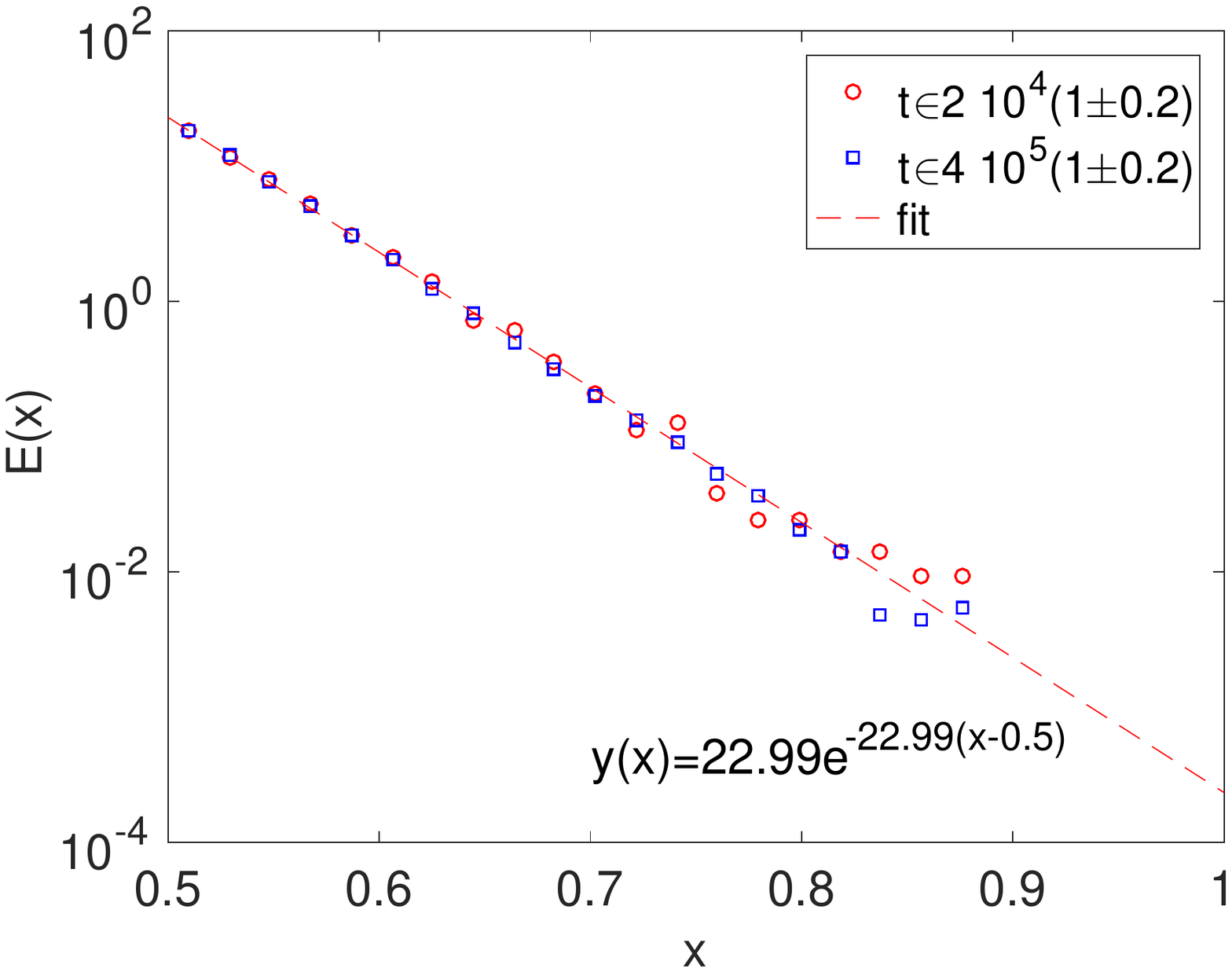}
\vspace{-2cm}
\caption{The empirical length distribution of `long' jumps associated with cage breakings 
is sampled in the two time intervals, $t \in10^4 [1.6,2.4]\tau$ and $ t\in10^5 [3.2,4.8]\tau$.
All data are fitted by the exponential function shown. The volume fraction of the system is $\phi=0.620$.
}
\label{fig:intermittency2}
\end{figure}

 To summarize, the phenomenology of glass formers and aging HSC is 
 experimentally~\cite{Weeks02,Courtland03,Cianci06,Li10,Valeriani11,Yunker09}
and numerically~\cite{ElMasri05,ElMasri10,Chaudhuri07,Pastore14}  well described, but
a unified   theoretical description~\cite{Ciamarra16,Pastore16,Robe16}  is not yet available.
RD has been proposed~\cite{Boettcher11,Robe16,Boettcher18} as a viable candidate 
and its validity is investigated  in  the present  work by  extensive MD simulations 
of three dimensional HSC which extend  over 6 order of magnitude in time.
We first provide a macroscopic characterization of the dynamics in terms of particle MSD 
and potential energy, and compare with homologous results~\cite{ElMasri10}.
We then proceed to investigate single particle jump statistics, and find a quantitative measure of the cage size  
  and the length distribution associated with  cage-breaking jumps. The information is used to identify quakes
  and to  show that they obey log-Poisson statistics, which is the  main prediction of  RD. The results, depicted in Fig.\ref{fig:quake_stat}
  buttress  Record Dynamics as a good  coarse-grained description of aging HSC systems.

 \section{Computation details and notation}
 
 We perform simulations of essentially hard sphere colloidal particles
using the model of Voightmann et al. \citep{voigtmann2004tagged}.
The interaction between colloids is modelled by the steeply repulsive potential

\[
U(r_{ij})=\frac{\epsilon}{3}\left(\frac{\sigma_{ij}}{r_{ij}}\right)^{36},
\]
where we take $k_{B}=1$ such that  $\epsilon$ is our unit
of both energy and temperature. The distance between a pair of particles
is denoted $r_{ij}=|{\bf r}_{i}-{\bf r}_{j}|$, while the length scale
of the repulsive potential between them is $\sigma_{ij}=(\sigma_{i}+\sigma_{j})/2$,
where $\sigma_{i}$ denotes the diameter of the $i$'th particle.
To avoid crystallization, we choose diameters from an uniform distribution
$\sigma_{i}\in[\sigma-\Delta\sigma:\sigma+\Delta\sigma]$, where 
$\sigma$ is our unit of length. Finally, all colloidal particles
have a mass $m$, which we choose as our unit of mass. From the definition
of energy, length and mass, the formal definition of the simulation
unit of time is $\tau=\sigma\sqrt{\epsilon/m}$. 
 This is the characteristic time it takes
an isolated  particle  to move its own diameter with ballistic motion
at  thermal speed. 
We note that while
mapping most of our units (and hence results) to experimental data
is straight forward, mapping of time scales should not be done using
the formal definition, since the latter  has no
physical relevance for glassy colloids. Time should rather be mapped
by matching emergent dynamical properties such as the diffusion coefficients
observed in experiments and simulations. 

Colloidal simulations were performed in the NVT ensemble
 at temperature $T=\epsilon/3$ as in Refs. \citep{voigtmann2004tagged,ElMasri10}.  Our systems  comprised
$N=50000$ colloidal particles in a cubic box with periodic boundary
conditions. The system volume was determined based on the desired
target volume fractions $\phi$ resulting in box sizes larger than $34\sigma$. Hence we
do not expect any finite-size effects in our data. We choose $\Delta=0.2$
corresponding to a $11.5\%$ polydispersity index\citep{ElMasri10}, since we
observed crystallization when using $\Delta\sigma=0.1$ as in Ref.
\citep{voigtmann2004tagged}. Taking the polydispersity average, the
volume fraction is given by $\phi=\pi\rho\sigma^{3}[1+\Delta]/6$.


We have simulated volume fractions $\phi=0.5$, $0.575$, $0.590$, $0.605$,
$0.620$, $0.633$, $0.647$, and $0.662$. The glass transition is
expected at $\phi_c\approx0.620$, hence we have seen
dynamical behavior ranging from simple time homogeneous diffusion to 
aging dynamics. During the
simulations, we continously monitored the local orientational order
parameters\citep{steinhardt1983} to ensure the system did not spontaneously crystallize.
For  each glassy system we ran two statistically independent replicas up
to times in excess of $10^{6}\tau$ ($4\times10^{8}$ integration
steps). We choose velocity rescaling as a thermostat due to its computational
efficiency. Particle velocities were rescaled every $0.25\tau$, while the
linear and angular momenta of the whole system were  reset every $1000\tau$ to prevent
flying ice-cube effects\citep{harvey98}. The dynamics was
numerically integrated using velocity-verlet with time step $\Delta t=0.0025\tau$
using a customized version of the Large Atomic Molecular Massively
Parallel Simulator (LAMMPS)\citep{PlimptonLAMMPS95}. Each glassy system
required about $60$ days of continuous simulation time on a $24$ core compute
node~\citep{AbacusHardware}. The total computational effort of the
simulations reported here is approximately $36$ core years.

\subsection{System preparation}
In experimental colloidal systems such as that of Yunker et al.~\cite{Yunker09},
soft NIPA microgel particles  shrink in size under optical heating. 
The  particles rapidly swell
when the heating is turned off, and 
if the initial volume fraction is sufficiently high,
the resulting volume fraction then exceeds its  critical value.
In this way,  glassy dynamics  with a well-defined initial time
can be observed experimentally.

To mimic the preparation of experimental glassy colloidal systems,
we insert $N$ mono-disperse ($\sigma_{i}=\sigma$) colloids in the
simulation box at random positions, and minimize the energy to eliminate
particle overlaps. Each particle  has an integer tag $i=1,\dots,N$.
To quench the system, we assign a unique size given by $\sigma_{i}=\sigma+\Delta\sigma(2i/N-1)$
to each particle.
This prevents any statistical correlation between the spatial position
and size of the particles, and furthermore prevents system-to-system
variation due to different realizations of finite-sized samples taken
from the size distribution.

The quenched polydisperse configuration will have strong overlaps between
particles and cannot be used as initial state, since
the numerical integration would be unstable due
to excessively large forces. On the other hand, minimizing the energy could,
 in principle, lead to arbitrary large configurational re-organizations,
 which would blur the definition of the time elapsed from the initial quench.
 Hence inspired by the experimental procedure, we
run a short simulation with a Langevin thermostat with a very high friction
of $\Gamma=50m\tau^{-1}$ and  a numerical integrator that
maximally displaces a bead by $0.05\sigma$ during one time step. During
a very short simulation $2\tau$ or equivalently $800$ integration steps),
all overlaps are removed  and the state is concomitantly  thermalized to thetarget
temperature.
This thermalized post-quench system state defines age zero for
the subsequent data production run. The procedure just described is followed for 
all the volume fractions investigated.

\section{Systemic properties}
 The MSD and the potential energy vs. time are both
 systemic properties obtained by averaging observables over all particles.
 Specifically, the data shown are obtained as follows:  a set of logarithmically equidistant points
is placed on the time axis,
and, for each particle,  the  MSD or potential energy  values    falling in each of the corresponding  intervals are 
time averaged
and assigned to the midpoint of the interval. The values thus obtained  are then averaged over all particles,
and a final average is carried out over the outcomes of  two  independent simulations. These outcomes 
are  however already
practically indistinguishable at the resolution level of our figures.
 
 The same repulsive interaction and size poly-dispersity are used
 as  in Ref.~\cite{ElMasri10}, but 
  our systems contain  $50000$  rather than $500$ and $4000$ particles and 
  we follow them for one more decade of simulation time.
  Finally, as explained in the previous section,
  our system is not initially compressed as theirs. Instead,
   the particle sizes are   initially inflated to
  achieve the desired volume fractions.
  We note for clarity that our time $t$ is  the system age, which is denoted  by $t_{\rm w}$ in 
 Ref.~\cite{ElMasri10}, a symbol we here reserve for  expressions having two time arguments.
 
For several values of the volume fraction $\phi$, the evolution of the
mean-square displacement is  plotted vs. time in Fig.~\ref{fig:MSD} using a logarithmic abscissa.
The lowest volume fraction, $\phi=0.575$,
produces standard diffusive behavior, as shown explicitly in the insert. For all other volume fractions,
the MSD grows, after a short transient,  as the logarithm of time for more than four decades of simulation.
For similar results, see~\cite{Boettcher11,Robe18}.
\begin{equation}
{\rm MSD}(t) = D_{\rm ln}(\phi) \ln(t/\tau)
\end{equation}
where $\tau$ is the smallest time unit in the simulation.
In the following, times will always be measured in units of $\tau$, and the
symbol will be omitted from the notation.
 The logarithmic rate of diffusion $D_{\rm ln}(\phi)$ decreases monotonically with 
increasing  volume fraction  $\phi$, and its  values are well
 fitted with two free parameters by  the  function  $y(\phi)=0.833(\phi-0.603)^{-1} \; 10^{-3}$.
 With  two  parameters to  four data points,  the evidence  the above  expression  provides is only  anecdotal.
Nonetheless, the divergence  it features at $\phi_0 \approx 0.603$ correctly  detects   the
presence of an upper limit to the validity of the 
logarithmic diffusion regime.
Finally, even though the  MSD data shown in our Fig.~\ref{fig:MSD} are rather  well fitted by logarithms
the small deviations from the fit, best seen for $\phi=0.620$, have a systematic  character which we do not attempt to
explain.

Figure~\ref{fig:epot} shows the time decay of the potential energy per particle, averaged over all particles.
After a short initial transient $t\sim 20\tau$, the decay is seen to be  a linear function of the logarithm of time. 
The initial value of the potential energy  increases, as expected, with increasing $\phi$. 
Finally, the logarithmic rates of change of the potential energies are, in order of increasing 
$\phi$,  $-[5.8; 5.9; 6.3; 7.9]10^{-3}\epsilon$. The clear growing trend can be contrasted with
 the decreasing trend of the logarithmic `diffusion' coefficient $D_{\rm ln}(\phi)$. This implies that the physical
 effect of a particle rearrangment  grows with growing density.
 Taken together, our observations suggest logarithmic time as a natural variable
for describing the dynamical evolution of systemic properties of glassy systems.

 Qualitatively, the data in~\cite{ElMasri10} concur with ours as far as the age decay of
 the potential energy per particle is concerned. Note however that 
 two different types of fit of comparable quality are offered in~\cite{ElMasri10} for the asymptotic form of the decay, none of which is
 identical to our simple logarithmic decay.
 In  Fig. 4 of the same reference, the particle MSD is plotted vs. the time lag $\tau$ spent after waiting 
time  $t_{\rm w}$, which is the traditional choice in studies of aging dynamics\footnote{To avoid notational confusion,
 we stress that our  symbol $\tau$ stands for 
the simulation unit of time and not for  the lag time, except for one paragraph of the final Section.}.
  It is nevertheless clear from the
same figure that the  MSD grows logarithmically  as a function of $\tau$ if $\tau \gg t_{\rm w}$, i.e.
when time and lag time can be identified. Finally, the figure's inset shows that the
power-law fits of the MSD vs lag-time are based on data which only cover 
little more than one decade.  Since  the exponent of the sub-diffusive MSD growth is itself a function of the
age, see Fig. 5, the sub-diffusive behavior described in~\cite{ElMasri10} cannot be simply described as 
a  power-law.
 
\section{Displacement statistics}
 In this section, the `cage size' is extracted from the central part of the Van Hove function,  
 the age dependence of the exponential tails is analyzed
 and   the length distribution of the associated  `long jumps' is shown to 
 be age independent.

The distribution of single particle displacements, aka self part of the
Van Hove function,  was investigated in Ref.~\cite{ElMasri10} for a number of
time lags after a fixed waiting time $t_{\rm w}$. The results, shown in their Fig. 6,
can be summarized as follows: The distribution has a central Gaussian part of zero mean, 
whose  a shape only depends weakly on the lag time. Large displacements of both 
signs are described by exponential tails, whose weight increases with increasing lag time. 

Our analysis is patterned on the   method introduced in  Ref.~\cite{Sibani05a} to describe   heat transfer in a spin-glass model
  and its results  concur broadly with those of Ref.~\cite{ElMasri10}
  and precisely with those of a more recent simulational study of 2D HSC~\cite{Robe18} which  
obtains a data collapse by scaling  the lag time with the system age, as we presently do.

The probability density function (PDF) of displacements
$\Delta x$ occurring over a  short time interval $\Delta t$ (lag time) for given values of the system age $t_{\rm w}$  is written as
   \begin{equation}
 G_{\Delta x|t_{\rm w},\Delta t}(x)=\sum_i \delta(x_i(t_{\rm w})-x_i(t_{\rm w}+\Delta t)- x),
 \end{equation}
 which  is normalized over all  values of the dummy variable $x$.
  Specifically, $G$ is sampled by collecting
 all  positional changes  $x_i(t_{\rm w}-\Delta t)-x_i(t_{\rm w})$
 occurring at age $t_{\rm w}$ over   time  intervals
 of duration  $\Delta t$, 
 i.e. in the  
  interval  $I(t_{\rm w})=( t_{\rm w}, t_{\rm w} + \Delta t )$.
  
 To improve the  statistics,  spherical and reflection symmetry is used  to 
 \emph{i)} 
merge  the independent displacements  in  the three orthogonal directions $x, y$ and $z$
 into  a single file, representing a fictitious  `x' direction, and  \emph{ii)} to  invert the sign of all negative displacements.
  Compatibly with the requirement $ \Delta t \ll t_{\rm w}$ needed
  to associate $G$  with   a definite  age $t_{\rm w}$, 
  a  $\Delta t$   much larger than the mean time between collisions
   is preferable, as it  accommodates  many in-cage rattlings.
 We note in passing  that in
  the opposite limit the particles move in near ballistic fashion and that  their displacements inherit  the Gaussian
 distribution  of the components of their velocities.

Figure~\ref{fig:intermittency} depicts  PDFs   of single particle displacements   in  a colloid of volume fraction $\phi=0.620$. 
Small displacements have an age independent  Gaussian PDF, corresponding to the staggered line,
while larger displacements strongly deviate from Gaussian behavior, as seen in
Fig.~6 of Ref.~\cite{ElMasri10}.           
The displacements occur  over short time intervals of length $\Delta t \ll t_{\rm w}$ and are
 sampled in three longer    observation  intervals of the form $[t, 1.1  t ]$ where time values  
   $t=2\; 10^4\tau,\; t=8\;10^4\tau\; $ and $t=8\; 10^5\tau$ were chosen. (We recall
   that $\tau$ is  our simulational time unit.)
   Since the length of these intervals is only a tenth of the time
   at which observations commence,
 aging effects  occurring during observation    can be neglected and  
   $t$  can be identified with the system age, i.e. 
   $t_{\rm w}\approx t$. In the left hand panel  of Fig.~\ref{fig:intermittency}  a single value  $\Delta t=100\tau$ was used
   for all data sampling, while in the right hand panel  values  proportional to the system age,
   $\Delta t=100,400$ and $4000\tau$ were used.

 That the central part of $G_{\Delta x|t_{\rm w},\Delta t}$ 
 is a  Gaussian distribution with zero mean indicates that  displacements of
 small length arise from  many independent and randomly oriented contributions,
 which stem  from  multiple in-cage rattlings. The typical size of the cage
 can then be identified with  the standard deviation of the Gaussian part of the PDF
which  is seen to be  independent of age. The Gaussian standard deviation of the data
is estimated to be $\sigma_G=0.05\sigma$ and the 
  `cage width' $c_{\rm w}$ defined as  the standard deviation of the three dimensional Gaussian displacement PDF is
 $c_{\rm w}=\surd 3 \sigma_G=0.086\sigma\approx 0.1\sigma$.
 Recall that  $\sigma$ is the average particle diameter.
 This result concurs with the estimate obtained from 2D simulations~\cite{Robe18}, which
 suggest "$\ldots$ a caging length between about 1\% and 10\% of
 a particle diameter.".

The  exponential tail is then produced by 
cage-breakings, i.e. displacements well beyond the  cage size.
The weight of the non-Gaussian tail seen in the left hand panel of Fig.~\ref{fig:intermittency}
is seen to decrease with increasing age while 
the length distribution of displacements of length exceeding $0.5\sigma_G$ is seen
in Fig.~\ref{fig:intermittency2} to be exponential and 
 age independent.
The right hand panel of Fig.~\ref{fig:intermittency} shows  that scaling 
 $\Delta t$ with the age $t_{\rm w}$ reasonably collapses the data.
 The same effect  is obtained in \cite{Robe18} for 2D colloidal suspensions and 
 for other values of the ratio $\Delta t/t_{\rm w}$.

\begin{figure*}[t!]
\vspace{-3cm}
$
\begin{array}{lr}
\includegraphics[width=.45\linewidth]{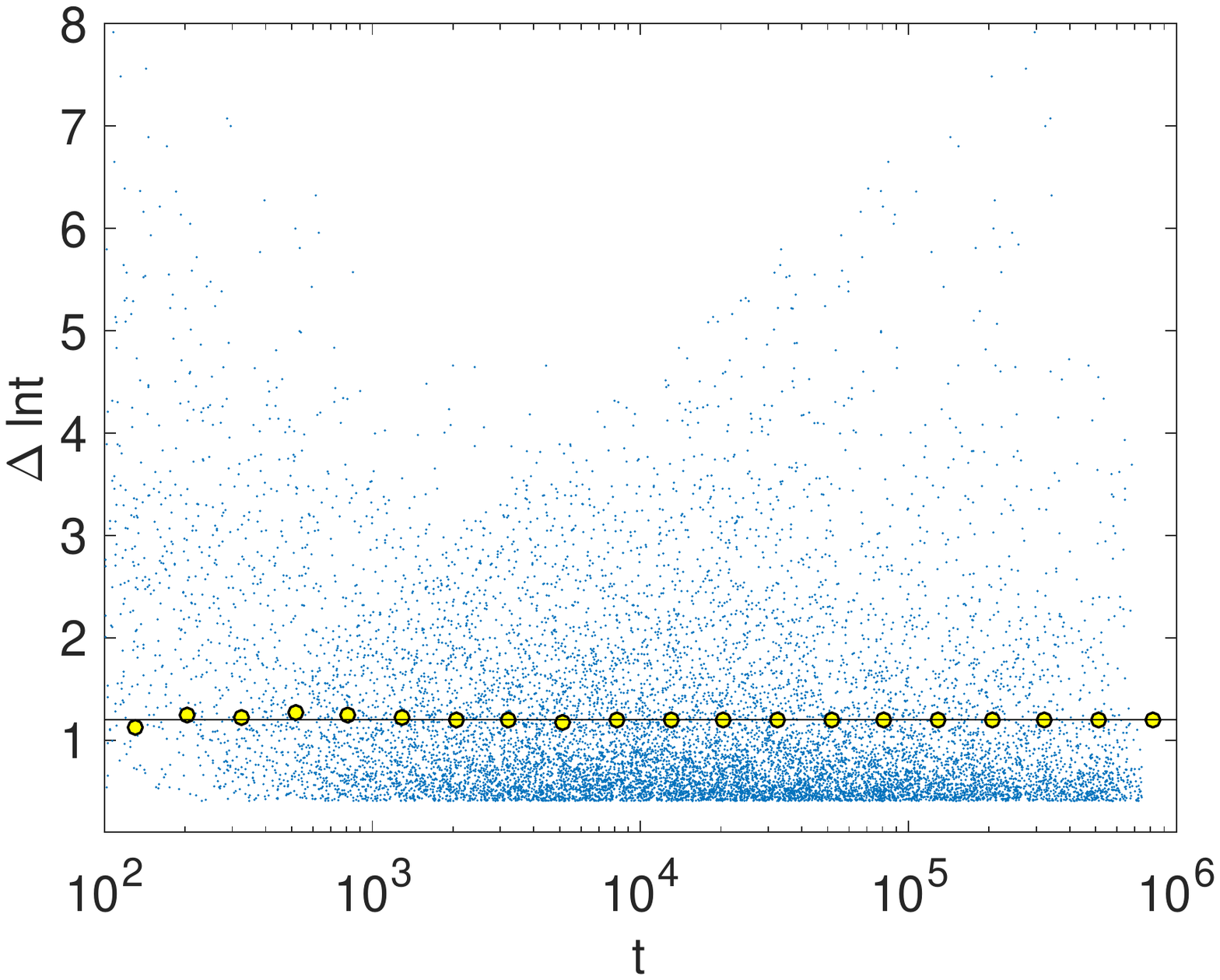} &
\includegraphics[width=.45\linewidth]{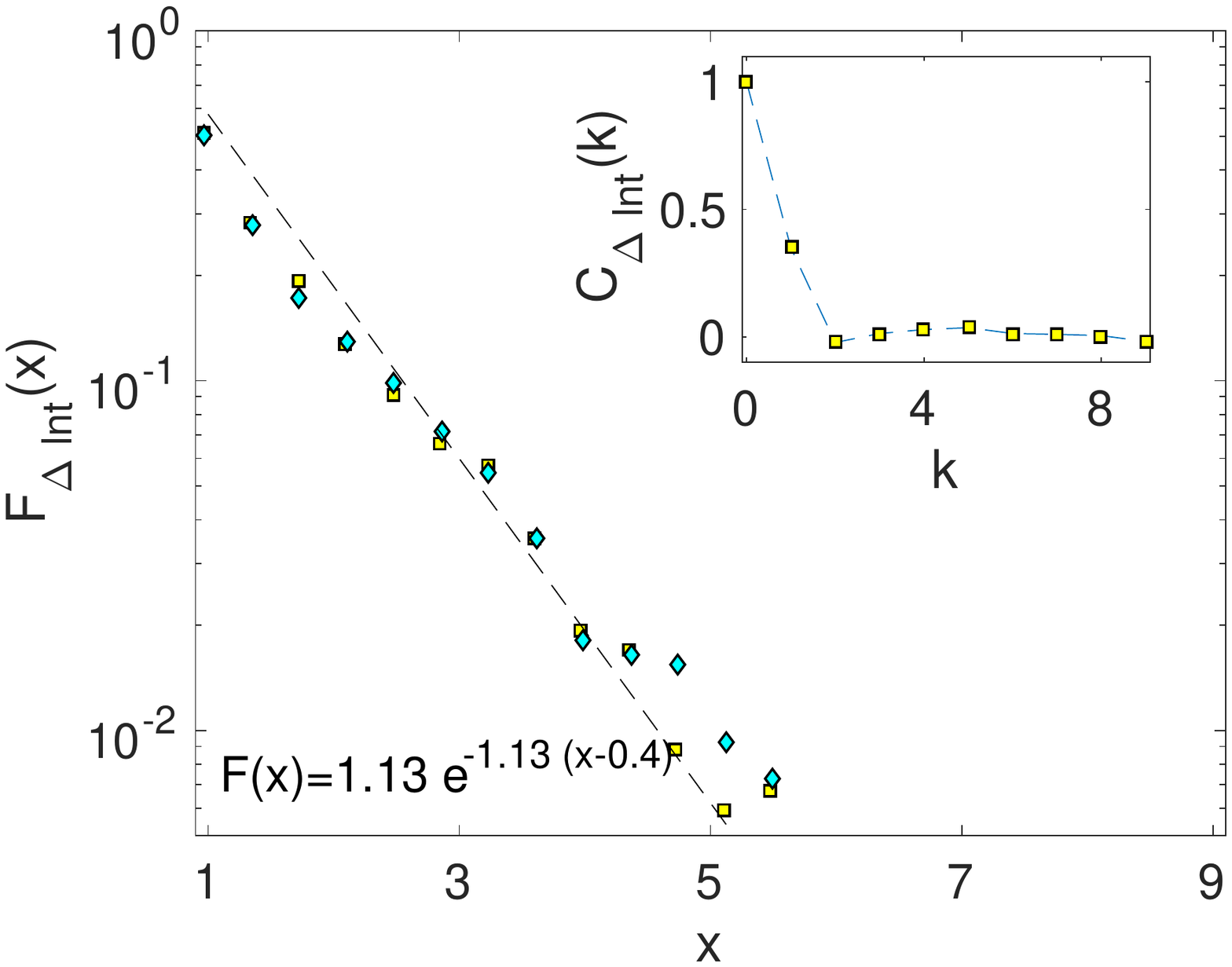}\\
\vspace{-0cm}
\end{array}
$
\vspace{-2cm}
\caption{Left hand panel: 
 Logarithmic waiting times  $\Delta{\rm \ln t}_k=\ln t_k/t_{k-1}$, where $t_k$ is the time of the $k$'th cage breaking
are observed in small domains of the simulation
box and  plotted, on a log scale,  vs. the time at which they are observed.
Only data with $\Delta{\rm \ln t}_k>0.4$ are included.
 The yellow circles are local binned log-time averages of these data and 
 the line is the  average logarithmic waiting time. That local averages are nearly  independent of log-time
 is evidence that the transformation $t\rightarrow \ln(t)$ renders the dynamics log-time homogeneous.
Right hand panel: the  PDF of the `log waiting times is  estimated and plotted for two independent sets of  simulational raw data,
using  yellow square and cyan diamond symbols, respectively. The  line is an exponential fit to both estimated PDFs.
The insert shows  the normalized autocorrelation function of the sequence of logarithmic waiting times corresponding to the 
yellow square PDF.
To a good approximation, the log-waiting times are uncorrelated and their PDF decays exponentially, which implies  that quaking is a log-Poisson process.
The volume fraction of the system is $\phi=0.620$.
}
\label{fig:quake_stat}
\end{figure*}

RD explains the observed   scaling  behavior by assuming  that the non-Gaussian tail
 of $G$ stems from quakes, which, as shown later, are log-Poisson distributed. Taking that for granted,
 the  average number of   quakes occurring  in an arbitrary  time interval $(t_1,t_2)$ is
  proportional to $\ln(t_2/t_1) $ and, in our case,  proportional to $\ln (1+\frac{\Delta t}{t_{\rm w}})\approx \frac{\Delta t}{t_{\rm w}}$.
 Replacing   the (small) 
 cross-over region between Gaussian and intermittent fluctuations by
 a sharp boundary, we   let $G(x)$ be a  truncated Gaussian pdf,  normalized  
 within  the cage, and  let $E(x)$ be  the pdf of the intermittent
 events, normalized in the semi-infinite interval outside the cage limit.
 With this notation,
the probability density  for an event of size $x$ 
 is given by 
 \begin{equation}
 G_{\Delta x | t_{\rm w},\Delta t}(x) = \left( G(x) +  \frac{\Delta t}{t_{\rm w}} E(x)\right) \left(1+  \frac{\Delta t}{t_{\rm w}}\right)^{-1}.
 \label{basic}
 \end{equation} 
 Choosing, as we did, $\Delta t=c t_{\rm w}, \; c\ll 1$  leads to 
  \begin{equation}
G_{\Delta x | t_{\rm w},\Delta t} (x) \approx (1- c) G(x) +  c E(x) \approx G(x) +  c E(x).
 \label{basic2}
 \end{equation}
 Since, as confirmed below, $E(x)$ is  independent of $t_{\rm w}$, 
 the observed data collapse follows from our choice of $\Delta t \propto t_{\rm w}$.
  
 The empirical distribution $E(x)$ of the length of the  `long' jumps associated to cage breakings,
 is sampled in two widely separated  time intervals and depicted in  Fig.~\ref{fig:intermittency2} 
  for  volume fraction $\phi=0.620$
 together with a one parameter data fit. The latter shows that  $E(x)$ is exponential and effectively
 age independent,
 as already assumed Eq.~\eqref{basic2}. The empirical standard deviation of  the jump length
 is in this case $\sigma_q=0.545\sigma$,  where $\sigma$ is the average particle diameter.
 Note that  $\sigma_q$ is an order of magnitude larger than the
  standard deviation  $\sigma_G$ describing  in-cage rattling. 
 The other  investigated volume fractions, $\phi=0.633,\; 0.647$ and $\phi=0.662$,
  show the same overall behavior and
 the corresponding jump length 
 standard deviations, $ \sigma_q=0.544, 0.543$ and $0.541\sigma$,
 show a small but systematic decrease  with increasing volume fraction. 
 
 Summarizing, our analysis of the self part of the Van Hove function indicates that cage-breakings
 occur at a rate which decreases with the inverse of the system age, see \cite{Robe16} and  \cite{Robe18}  for
 corroborating experimental and simulational evidence. Consequently,
the  number of these events increases logarithmically with time. Since cage-rattlings do not contribute
 to the  particle diffusion, the result explains  the observed logarithmic diffusion.
 The next section provides fuller evidence for RD and for the logarithmic nature of HSC diffusion.
\section{Quake statistics and Record Dynamics}
That  the rate of irreversible rearrangements in HSC decreases with the inverse of the system 
age  was  noticed by 
~\cite{Robe16} in  previously published  experimental 2D data~\cite{Yunker09} and 
was later confirmed by  2D MD simulations~\cite{Robe18}. 
The  observation
clashes~\cite{Boettcher18} with widespread CTRW model predictions and  
 strengthens   a competing RD description of the mechanism beyond aging in HSC and other complex systems.
 
We provide detailed statistical evidence in favor of RD, by extracting irreversible dynamical events
called quakes, which are statistically independent, homogeneously distributed on a logarithmic time axis and
described by an exponential distribution of the `logarithmic waiting times' $\ln t_k -\ln t_{k-1}$, where $t_k$ is the time of the k'th quake.
Together, these properties  imply that quaking is a log-Poisson process, i.e. a Poisson process whose average has a logarithmic 
time dependence.
The key step of quake identification  in a given setting has some leeway,
but, as we argue below, in spatially extended systems spatio-temporal correlations  play a major role:
The existence of spatial domains, a property which reflects the strong spatial heterogeneity of glassy dynamics~\cite{Chaudhuri07},
 is required in  a RD description of aging systems.
 
 Spatially extended aging systems of size $N$~\cite{Sibani03} 
contain an extensive  number $\alpha(N)\propto N$ of 
 equivalent spatial domains such that  events occurring in different domains are statistically independent.
 Events occurring in the same domain have  long-lived  temporal correlations, which are
 formally removed in RD  by the global variable 
transformation $t \rightarrow \ln t$.  If this device works, 
 the total number of quakes occurring in the system between times $t_{\rm w}$ 
 and $t>t_{\rm w}$ is a Poisson
 process with average
 \begin{equation} 
 \mu(t,t_{\rm w})=\alpha(N) r_q  \ln(t/t_{\rm w}),
 \label{grandPoisson}
 \end{equation} where $r_q$ is the average logarithmic quake rate in each domain.
 The number of quakes  is extensive and grows  at a constant rate in log-time
 and at a rate proportional to $1/t$ in real time. Within each domain, the rate $r_q$
 can be read off the log-waiting time PDF $F_{\Delta {\ln t}}(x)=r_q e^{-r_q x}$, i.e. the probability density
 that the log-waiting time to the next quake equals $x$.

To ascertain  the applicability of  the  above RD scheme in a specific  system,
  the log-waiting times between successive quakes within each domain, $d\ln{t}=\ln t_k - \ln t_{k-1}$
 are formed and their statistical properties are checked: 
specifically,  log-waiting times must be 
 independent and  identically distributed stochastic variables uniformly
 distributed on a logarithmic time axis, as shown  in the left hand panel of
 Fig.~\ref{fig:quake_stat}.
\begin{figure}[t!]
\vspace{-2cm}
\includegraphics[width=.95\linewidth]{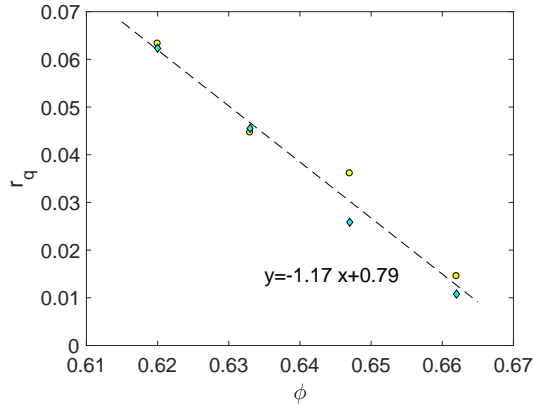} 
\vspace{-0cm}
\vspace{-2cm}
\caption{The logarithmic quake rate \emph{per particle}, $r_q$, is obtained from the decay coefficient  of the log-waiting time PDF,
aka the logarithmic quake rate per domain, 
(see main text) and plotted vs. the volume fraction $\phi$. The two values  given for each $\phi$ are obtained from two independent simulations.
Their average is well  fitted by the shown line.
%
%
}
\label{fig:quake_rate}
\end{figure}
Secondly, the PDF of the log-waiting time must be exponential, as shown in the right hand panel of the same figure.

 To see why   a domain structure
 is inherently part of the RD description of a  system
  of  $N$ degrees of freedom, assume for a moment that 
 domain partitioning can be  neglected.
 The time lags $\delta_k $ between successive quake times then decrease with
 increasing  $N$,
 and, since $d\ln t_k\stackrel{\rm def}{=}\ln( t_k/t_{k-1})=\ln(1+\delta_k/t_{k-1}) \approx \delta_k/t_{k-1}$,
  the `log waiting times' can never be   log-time homogeneous  in the limit $N\rightarrow \infty$.
The latter property   clearly requires $\delta_k \propto t_k$, which is
 incompatible with $\delta_k \propto N^{-1}$. 
 In contrast,  if the dynamics is   time translational 
 invariant,  $ \ln(t/t_{\rm w})$ in Eq.~\eqref{grandPoisson} is replaced by
 $t-t_{\rm w}$ and  the prefactor to $r_q$ is $N$, no matter how the system is partitioned.
 
Usually, only a minuscule fraction of  configuration space is explored
 during  an aging process, and 
  many  variables do not participate in any quake. Hence, the 
 domain  size  can grow in time 
 with no changes in  $\alpha(N)$, which is best 
 understood as the number of \emph{active} domains where quake activity occurs.
 
 A statistical analysis of domain size and domain growth is an important issue not presently considered.
In the present work, the simulation box is simply subdivided  into $16^3$ 
equal and adjacent sub-volumes, each  containing,  on average, slightly more than ten particles.
The size was chosen self-consistently as the largest possible size yielding log-time homogeneous quake statistics.
Many sub-volumes in the partition had no quaking activity, showing the 
strong spatial heterogeneity of the HSC dynamics.
	
\section{Summary  and discussion}

In this study,   large 3D poly-disperse  hard sphere colloidal  systems (HSC) are studied by extensive MD simulation.
The initial state is generated by a sudden expansion of the particles' volume, 
leading to volume fractions $\phi$ both below and above the critical value. The systems' development
is subsequently followed for more than six decades in time.
 
At the   systemic level our analysis concerns  the growth of the particles' mean square displacement (MSD)
and the decay of their potential energy as a function of time, both  shown to be simple 
functions of the logarithm of the system age above the critical density.
 At least for the MSD, this concurs with similar results, both experimental~\cite{Boettcher11}
and numerical~\cite{Robe18}, but  apparently differs from the more usual~\cite{ElMasri10} description in terms of  power-laws.

At the level of single particle displacement,
we study the self part of the Van Hove function and confirm~\cite{Chaudhuri07,ElMasri10} that 
cage rattlings and cage-breakings  have a Gaussian
and exponential length distribution, respectively. We find that both distributions are age independent,
and that the weight
of the exponential tail is a function of the ratio of the lag time and  the system age. This extends a result recently
obtained~\cite{Robe18} for a 2D HSC system and is akin to the behavior of heat transfer in a spin glass model~\cite{Sibani05a}.

We note that one-time averages, e.g.   the MSD and the potential energy per particle, are most naturally described
in term of a single time variable $t$, i.e. the time elapsed from the quench into the glassy phase.
Experimentally, the time origin can be difficult to determine, and a description in terms of two 
variables, conventionally age $t_{\rm w}$ and lag time $\tau \stackrel{\rm def}{=}t-t_{\rm w}$ is generally 
preferred.
(For notational convenience our  symbol  $\tau$ is here re-defined and  used 
for  the lag time).
We have shown that  MSD per particle has the form MSD$(t)=D_{\rm ln} \ln t= \ln(t_{\rm w}+\tau)$, which 
grows first linearly in $\tau$  
and than crosses over to a logarithmic growth. This is seen in Fig. 4 of Ref.~\cite{ElMasri10}, but the data are there  interpreted in terms of a several of power laws with 
 age dependent exponents, see Eq.(10) and Fig. 5 of the same reference.
 
 Even though a power-law with a  small exponent and  a logarithm may look similar over restricted time scales,
 the physical pictures behind them  are different~\cite{Sibani13}.  Dense HSC dynamics is  usually
  interpreted in   CTRW terms~\cite{Pastore14,Ciamarra16,ElMasri10},
 while logarithmic laws, or equivalently, the fact that macroscopic rates fall off as the inverse of the system age,
clearly  support  the  competing  RD description~\cite{Boettcher11,Sibani14,Robe16,Robe18}.
 To clearly distinguish between the two pictures  motivates our detailed
 analysis  of the statistics of quake events. The system is partitioned  into a number of spatial domains
  within which  the  quake statistics is shown to be log-Poissonian,  as
  expected from  RD.
 
Two related questions remain: what is the nature of the records which RD refers to, and how should domains be understood.
The logarithmic decay of the average potential energy per particle shown in Fig.~\ref{fig:epot} indicates that the
particle--or, equivalently, the voids between them--become more uniformly distributed in space.

Accordingly, it  becomes increasingly difficult for  random cage-rattling to create  a void large enough for  a cage-breaking to happen.
 In other words, the latter requires the crossing of a mainly entropic barrier, whose height
 increases with time. Records events  in exploring the associated hierarchy of free energy  barriers  generate  quakes,
 and the increase in the size of the barrier crossed corresponds to an increasing 
number of adjacent particles which must be re-arranged in the process, i.e. to a growing domain.
This type  of growing domains relate to the dynamical nature of rare local density fluctuations, 
and are not easily expressed in terms of single particle density fluctuations.
They are also seen  in 
 the aging dynamics of  the one-dimensional  kinetically constrained model
 known as the parking lot model~\cite{Sibani16}.  There,
    the key feature  is
 that random rearrangements of   a domain produce a quake
 in a time that grows strongly with the domain size, which is 
 also the central  hypothesis in the `cluster' model
 of Ref.~\cite{Boettcher11,Becker14}. By identifying  domains
   and studying their growth in HSC,  it should be possible to ascertain 
whether the same mesoscopic mechanism is at play in HSC systems.\\
\vspace{.25cm}

\noindent {\bf Acknowledgments.}
The first author has greatly  benefitted from numerous discussions with Stefan Boettcher.
Our computational effort was supported
by the DeiC National HPC Center, University of Southern Denmark.

\end{document}